\documentclass[amsmath,twocolumn]{aastex702}

\begin{document}

\title{Shock breakout from mildly relativistic ejecta in a dense wind: the case of EP260321a/SN~2026gzf}

\author[orcid=0000-0002-7043-6112,sname='A. Suzuki']{Akihiro Suzuki}
\affiliation{Research Center for the Early Universe, School of Science, University of Tokyo, Bunkyo-ku, Tokyo, 113-0033, Japan}
\email[show]{akihiro.suzuki@resceu.s.u-tokyo.ac.jp}  

\author[orcid=0000-0003-2611-7269,sname='K. Maeda']{Keiichi Maeda} 
\affiliation{Department of Astronomy, Kyoto University, Kitashirakawa-Oiwake-cho, Sakyo-ku, Kyoto 606-8502, Japan}
\email{}

\author[orcid=0000-0002-4060-5931,sname='T. Shigeyama']{Toshikazu Shigeyama}
\affiliation{Research Center for the Early Universe, School of Science, University of Tokyo, Bunkyo-ku, Tokyo, 113-0033, Japan}
\email{}


\begin{abstract}
We present shock breakout (SBO) modeling of the recently discovered X-ray transient EP260321a detected by the \textit{Einstein Probe} mission.
Our semi-analytic model, based on our previous work, follows the interaction between a supernova ejecta with a mildly relativistic outer envelope and a dense, wind-like circumstellar medium (CSM) by using a thin-shell approximation that incorporates relativistic effects. 
We find that the observed properties of the X-ray emission are well explained by the breakout emission powered by a high-velocity envelope with a kinetic energy of  $\sim3.5\times10^{49}\,\mathrm{erg}$ (excluding the supernova ejecta) and a dense wind characterized by a mass-loss rate of
$\dot{M}\simeq1.2\times 10^{-3}(v_\mathrm{w}/10^3\,\mathrm{km\,s}^{-1})\,M_\odot\,\mathrm{yr}^{-1}$,
where $v_\mathrm{w}$ is the wind velocity. 
The observed burst duration requires the dense CSM to extend up to $\sim350\,R_\odot$, corresponding to a CSM mass of $\sim 10^{-5}\,M_\odot$. 
These results demonstrate that SBO observations provide a sensitive probe of mass loss from stripped-envelope supernova progenitors shortly before core collapse.
\end{abstract}



\section{Introduction\label{sec:intro}}

Core-collapse supernovae (CCSNe), the terminal explosions of massive stars with initial masses above $\sim8$--$10\,M_\odot$, are among the most extensively studied transient phenomena in astrophysics. 
Nevertheless, their behavior during the earliest phases of the explosion remains poorly understood. 
The collapse of the iron core releases a large amount of gravitational energy in the form of neutrinos, a fraction of which is deposited into the surrounding material and revives the stalled shock. 
The resulting shock wave propagates through the stellar envelope and ultimately disrupts the star. 
The first electromagnetic signal from the explosion is produced when the shock reaches the surface of the progenitor star, giving rise to the supernova shock breakout (SBO) \citep{1973ApJ...180L..65F,1974ApJ...187..333C,1978ApJ...225L.133F,1978ApJ...223L.109K}. 
Because SBO emission directly probes the immediate environment of the explosion, it provides unique information on the progenitor structure and the final stages of stellar evolution that is otherwise difficult to obtain \citep[see][]{2017hsn..book..967W}.

Despite its importance, only a limited number of SBO events have been reported to date, primarily because of their intrinsically short durations. 
In particular, in the X-ray regime, which is expected to probe SBOs from compact stripped-envelope progenitors, SN~2008D remains the only event unambiguously identified as an SBO \citep{2008Natur.453..469S,2008Sci...321.1185M,2009ApJ...702..226M}. 
Several low-luminosity gamma-ray bursts (llGRBs), such as GRB~060218/SN~2006aj, have also been suggested to originate from SBOs \citep{2006Natur.442.1008C,2006Natur.442.1014S,2007ApJ...667..351W}, although their interpretation remains under debate \citep[e.g.,][]{2007ApJ...659.1420T,2016MNRAS.460.1680I,2025MNRAS.542.1269I}.

Wide-field X-ray survey missions with accurate localization are therefore essential for detecting and characterizing such fast X-ray transients. 
Recently, the {\it Einstein Probe} mission \citep{2022hxga.book...86Y} reported the discovery of the fast X-ray transient EP260321a \citep{2026arXiv260610014Y}, which shares several properties with previously reported SBO candidates, including SN~2008D. 
Extensive follow-up observations across the electromagnetic spectrum successfully identified a nearby counterpart at a redshift of $z\simeq 0.034$ \citep{2026arXiv260610009C,2026arXiv260610002M,2026arXiv260609992O,2026arXiv260610011R}. 
The optical transient was classified as the broad-lined Type Ic supernova (Ic-BL SN) SN~2026gzf, exhibiting spectroscopic properties similar to those of Ic-BL SNe associated with GRBs. 
Unlike previously reported Ic-BL SNe accompanied by llGRBs, however, the absence of bright radio emission disfavors the presence of a powerful relativistic jet that would produce luminous synchrotron radiation \citep{2026arXiv260610002M,2026arXiv260609992O,2026arXiv260610011R}.

One of the most intriguing aspects of this event is the reported detection of precursor activity more than 10 years before the X-ray transient \citep{2026arXiv260610009C}. 
If physically associated with EP260321a/SN~2026gzf, this optical precursor would provide compelling evidence for  mass ejection shortly before the terminal explosion. 
In addition, EP260321a exhibited an unusually long X-ray duration, exceeding $10^3\,\mathrm{s}$, significantly longer than the timescale expected for SBO emission emerging directly from the surface of a compact Wolf--Rayet progenitor. 
These observational properties instead favor a wind breakout scenario, in which the SBO emission is produced when the supernova shock propagates through an optically thick circumstellar medium (CSM) surrounding the progenitor. 
Indeed, such wind SBO models have been proposed to explain SN~2008D and GRB~060218/SN~2006aj \citep[e.g.,][]{2012ApJ...757..178G,2014ApJ...780...18G,2012ApJ...759..108S,2014ApJ...788..113S,2014ApJ...788L..14S,2018ApJ...853...52O}.

In this work, we apply our SBO model, originally developed to explain llGRBs \citep{2017ApJ...834...32S,2019ApJ...870...38S}, to the newly discovered event EP260321a/SN~2026gzf. 
By modeling the prompt X-ray light curve, we constrain the properties of the dense CSM and the embedded supernova ejecta responsible for the breakout emission. 
Our analysis provides insight into the poorly understood origin of dense CSM around stripped-envelope supernova progenitors and the physical mechanism responsible for highly energetic explosions in these systems.

This paper is organized as follows. 
In Section~\ref{sec:model}, we briefly describe our SBO model. 
In Section~\ref{sec:result}, we present the results of the X-ray light-curve modeling of EP260321a. 
In Section~\ref{sec:discussion}, we discuss the implications of the inferred physical properties and place EP260321a in the broader context of SBO and llGRB studies. 
Finally, we summarize this paper in Section~\ref{sec:summary}. 

\section{Shock Breakout Model\label{sec:model}}

In this section, we describe the SBO model adopted in this work. Since the formulation is largely based on our previous studies \citep{2017ApJ...834...32S,2019ApJ...870...38S}, we provide only a brief summary of the model assumptions, numerical setup, and parameters relevant to the present analysis.

\subsection{SN ejecta and circumstellar medium}

We consider the interaction between freely expanding SN ejecta and a CSM under the assumption of spherical symmetry, together with the associated production and escape of radiation from the shocked region.

The ejecta is assumed to be in homologous expansion. Under this assumption, the velocity of a fluid element located at radius $r$ at time $t$ is given by $\beta = r/(ct)$, where $\beta$ is the velocity normalized by the speed of light $c$. The corresponding Lorentz factor is $\Gamma = (1-\beta^2)^{-1/2}$. 

The ejecta density profile is described by a broken power law in the four-velocity $\Gamma\beta$,
\begin{equation} 
\rho_\mathrm{ej}(t,r)=\left\{ \begin{array}{l} \rho_0\left(\tfrac{t}{t_0}\right)^{-3}(\Gamma_\mathrm{br}\beta_\mathrm{br})^{-n}\ \mathrm{for}\ \Gamma\beta\leq \Gamma_\mathrm{br}\beta_\mathrm{br}, 
\\ \rho_0\left(\tfrac{t}{t_0}\right)^{-3}(\Gamma\beta)^{-n} 
\\ \hspace{2em}\mathrm{for}\ \Gamma_\mathrm{br}\beta_\mathrm{br}<\Gamma\beta<\Gamma_\mathrm{max}\beta_\mathrm{max}, \end{array} \right. 
\label{eq:density} 
\end{equation}
where $\Gamma_\mathrm{br}\beta_\mathrm{br}$ denotes the break four-velocity separating the inner flat component and the outer power-law component. Throughout this work, we fix the break velocity at $\Gamma_\mathrm{br}\beta_\mathrm{br}=0.1$. 
Optical spectra obtained within a few days of the explosion consistently show ejecta velocities reaching $\sim 0.1c$ \citep{2026arXiv260610002M,2026arXiv260609992O,2026arXiv260610011R}. 
These observations strongly suggest that a high-velocity ejecta component extending beyond $\sim 0.1c$ plays a key role in producing the SBO emission. 
Although a substantially more massive ejecta component is expected at lower velocities, its contribution to the SBO signal is likely negligible. 
Therefore, for the purpose of modeling the SBO emission, we consider only the high-velocity outer envelope of the ejecta.
The maximum velocity $\beta_\mathrm{max}$ defines the outer edge of the ejecta, initially located at $r=c\beta_\mathrm{max}t_0$ at the initial time $t=t_0=10\,\mathrm{s}$. 
The corresponding Lorentz factor, $\Gamma_\mathrm{max}=(1-\beta_\mathrm{max}^2)^{-1/2}$, is treated as a free parameter.

In the present study, we adopt an outer density slope of $n=5$. High-velocity ejecta with such a shallow density profile have frequently been invoked in early spectral modeling of energetic supernovae \citep{2019Natur.565..324I,2023MNRAS.522.2267M}. Moreover, multidimensional simulations of engine-driven and jet-driven explosions consistently produce high-velocity ejecta characterized by $\rho_\mathrm{ej}\propto r^{-5}$ or $r^{-6}$ in the outermost layers \citep[e.g.,][]{2019ApJ...880..150S,2022ApJ...925..148S,2024PASJ...76..863S}, suggesting efficient energy deposition into the outer ejecta.

The density profile in Equation~(\ref{eq:density}) is normalized by $\rho_0$. Rather than using $\rho_0$ directly, we parameterize the model by the initial kinetic energy contained in the ejecta described by Equation~(\ref{eq:density}),
\begin{equation}
E_\mathrm{rel}
=
4\pi c^5
\int_{0}^{\beta_\mathrm{max}}
\Gamma(\Gamma-1)
\rho_\mathrm{ej}(t_0,r)
\beta^2
\mathrm{d}\beta,
\end{equation}
which is treated as a free parameter in the following analysis\footnote{We note that this definition has changed from \cite{2019ApJ...870...38S} who employed a lower bound $\beta=1/\sqrt{2}.$}.
For the employed density structure, inner layers below $\beta<\beta_\mathrm{br}$ subdominantly contribute to the kinetic energy. 
Also, the kinetic energy of the more massive SN ejecta supposedly embedded below $0.1c$ is not taken into account in $E_\mathrm{rel}$. 

We assume the following CSM density structure inversely proportional to the radius squared,
\begin{equation}
\rho_\mathrm{w}(r)=Ar^{-2},
\qquad
(r<R_\mathrm{out}),
\end{equation}
where the density normalization is expressed as
\begin{equation}
A=
5\times10^{11}
A_\ast\,
\mathrm{g\,cm^{-1}}.
\end{equation}
The CSM extends to an outer radius $R_\mathrm{out}$, beyond which the density is truncated.

The inverse-square density profile corresponds to steady mass loss at a constant wind velocity. For a (non-dimensional) CSM density parameter of $A_\ast=1$ and a wind velocity of $10^3\,\mathrm{km\,s^{-1}}$, the corresponding mass-loss rate is approximately $10^{-5}\,M_\odot\,\mathrm{yr^{-1}}$, typical of Wolf--Rayet stars. The total CSM mass enclosed within $R_\mathrm{out}$ is
\begin{equation}
M_\mathrm{CSM}
\simeq
4\pi A R_\mathrm{out}
\simeq
10^{-5}M_\odot
\left(
\frac{A_\ast}{100}
\right)
\left(
\frac{R_\mathrm{out}}{500\,R_\odot}
\right).
\end{equation}

Motivated by the long duration of EP260321a, we focus on the regime in which the CSM is sufficiently dense that the photosphere resides within the CSM rather than at the progenitor surface. Assuming a constant opacity $\kappa$, the optical depth of the material from $r$ to $R_\mathrm{out}$ is
\begin{equation}
\tau_\mathrm{w}(r)
=
\int_r^{R_\mathrm{out}}
\kappa
\rho_\mathrm{w}(r')
\mathrm{d}r'
=
\kappa A
\left(
\frac{1}{r}
-
\frac{1}{R_\mathrm{out}}
\right).
\end{equation}
The photospheric radius, defined by $\tau_\mathrm{w}(R_\mathrm{ph})=1$, is therefore given by
\begin{equation}
R_\mathrm{ph}
=
\left(
\frac{1}{\kappa A}
+
\frac{1}{R_\mathrm{out}}
\right)^{-1}.
\end{equation}
For $R_\mathrm{out}\gg R_\mathrm{ph}$, this expression can be approximated as
\begin{equation}
R_\mathrm{ph}
\simeq
\kappa A
=
10^{13}\,\mathrm{cm}
\left(
\frac{\kappa}{0.2\,\mathrm{cm^2\,g^{-1}}}
\right)
\left(
\frac{A_\ast}{100}
\right).
\label{eq:photosphere}
\end{equation}

\subsection{Thin-shell approximation}

At the initial time $t_0=10\,\mathrm{s}$, the outermost ejecta and the inner edge of the CSM are assumed to be in contact at $r=c\beta_\mathrm{max}t_0$. 
At later times, the ejecta interacts with the surrounding CSM, generating forward and reverse shocks that convert kinetic energy into thermal energy.

To describe the subsequent evolution, we adopt the thin-shell approximation developed by \citet{2017ApJ...834...32S}. 
In this approach, the shocked ejecta and shocked CSM are assumed to be confined within a geometrically thin shell bounded by the forward and reverse shocks. 
The evolution of the shell is then determined by solving the conservation equations for mass and momentum.

In both the non-relativistic and ultra-relativistic limits, this treatment reproduces the corresponding self-similar solutions of \citet{1982ApJ...258..790C} and \citet{2006ApJ...645..431N}, respectively. 
Furthermore, \citet{2017ApJ...834...32S} demonstrated that the thin-shell approximation reproduces the shell evolution obtained from relativistic hydrodynamic simulations for a relevant range of parameter space.

\subsection{Radiative diffusion and light curve calculation}

The shocked shell stores internal energy generated primarily at the reverse shock, where the kinetic energy of the incoming ejecta is dissipated and converted into radiation. 
In addition to the dynamical evolution, our model follows the evolution of this radiation energy while accounting for radiative losses from the shell.

The emergent luminosity is estimated using the diffusion approximation. 
We assume that the dissipated shock energy is immediately transferred to radiation as long as the shell is optically thick. 
The radiation in the shell then diffuse through the shell at the diffusion velocity corresponding to the shell optical depth. 
Throughout this work, we calculate the photon diffusion velocity for an effective opacity of $\kappa_\mathrm{eff}=0.2\,\mathrm{cm^2\,g^{-1}}$, corresponding to electron scattering opacity for fully ionized, hydrogen-free material. 
The optical thickness $\tau_\mathrm{sh}$ of the shell is then defined as
\begin{equation}
    \tau_\mathrm{sh}=\frac{\kappa_\mathrm{eff}M_\mathrm{sh}}{4\pi R_\mathrm{sh}^2},
\end{equation}
for the shell mass $M_\mathrm{sh}$ and the radius $R_\mathrm{sh}$. 
The internal energy supply at the forward and reverse shock fronts is switched off after the shell optical thickness drops down below unity, $\tau_\mathrm{sh}<1$. 
After this optically thick-to-thin transition, the luminosity of the shell rapidly decreases. 
The energy loss rate through radiative diffusion can be computed from the following formula taking into account relativistic effect,
\begin{equation}
    \dot{E}(t)=4\pi R_\mathrm{sh}^2u_\mathrm{sh}v_\mathrm{diff},
\end{equation}
where $u_\mathrm{sh}$ is the average radiation energy density within the shell and the diffusion velocity $v_\mathrm{diff}$ is computed as follows,
\begin{equation}
    v_\mathrm{diff}=\mathrm{min}\left[\frac{c(1-\beta_\mathrm{sh}^2)}{(3+\beta_\mathrm{sh}^2)\tau_\mathrm{sh}+2\beta_\mathrm{sh}},c(1-\beta_\mathrm{sh})\right],
\end{equation}
with $\beta_\mathrm{sh}$ being the shell velocity normalized by $c$. 

Finally, the observable light curve is obtained by taking into account light-travel-time effects across the emitting surface. 
Following \citet{2019ApJ...870...38S}, the observed luminosity at observer time $t_\mathrm{obs}$ is calculated by integrating the local emission over equal-arrival-time surfaces. 
In practice, for a radial shell with radius $R_\mathrm{s}(t)$, velocity $\beta_\mathrm{s}(t)$ (corresponding Lorentz factor of $\Gamma_\mathrm{sh}$), and radiative energy loss rate $\dot{E}(t)$ evolving with time $t$, we use the following formula,
\begin{equation}
    L_\mathrm{bol}(t_\mathrm{obs})=c\int 
    \frac{\dot{E}(t)}{R_\mathrm{sh}(t)\Gamma_\mathrm{sh}^3[1-\mu \beta_\mathrm{sh}(t)]^3}dt,
\end{equation}
with $\mu$ being the direction cosine of the radial shell with respect to the line of sight;
\begin{equation}
    \mu=\frac{1}{R_\mathrm{sh}(t)}(t-t_\mathrm{obs}),
\end{equation}
as long as $\beta_\mathrm{sh}(t)\leq \mu\leq 1$ is satisfied (see, Appendix of \citealt{2019ApJ...870...38S}, for detail). 
This correction becomes particularly important when the emitting shell expands at mildly relativistic velocities.

\section{Results\label{sec:result}}
\begin{figure}[tbp]
\includegraphics[scale=0.64]{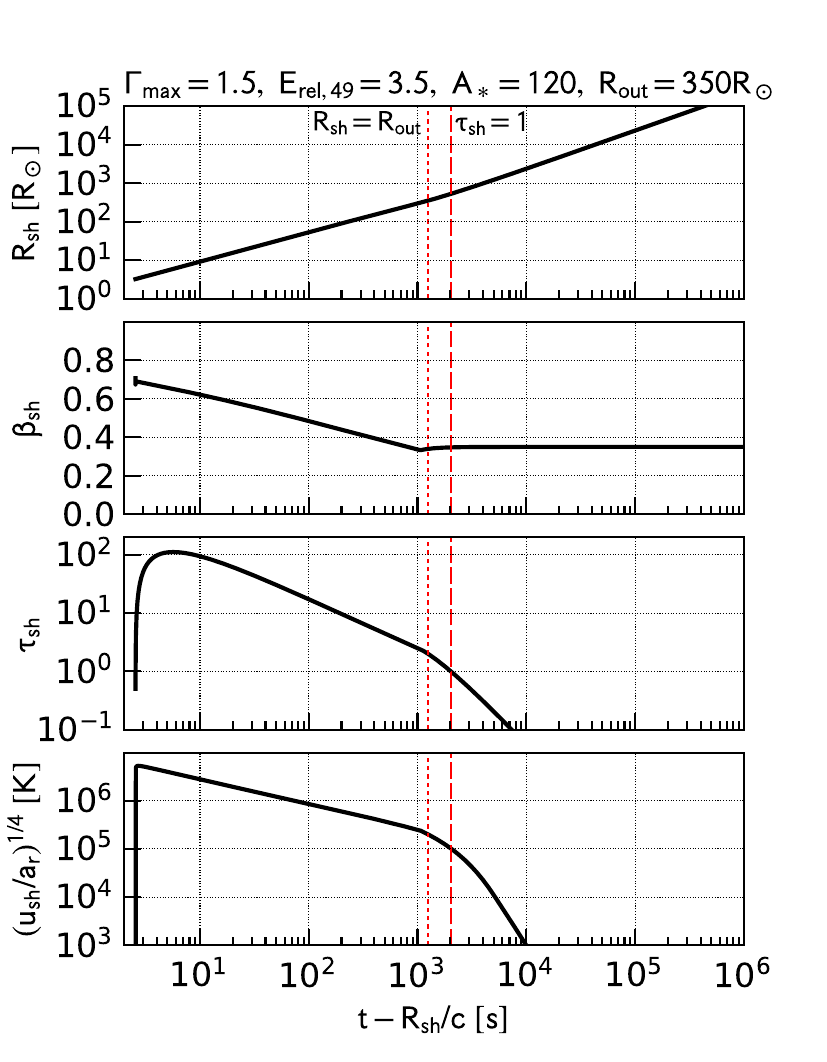}
\caption{Temporal evolution of the shell properties. 
The shell radius, the velocity, the optical thickness, and the equilibrium temperature are plotted as a function of the delay time $t-R_\mathrm{sh}/c$ from top to bottom. 
In each panel, two characteristic epochs,  corresponding to the optically thick-to-thin transition ($\tau_\mathrm{sh}=1$) and the shock emergence from the CSM outer radius ($R_\mathrm{sh}=R_\mathrm{out}$), are indicated by the vertical lines (dashed and dotted). }
\label{fig:dynamics}
\end{figure}
\subsection{Dynamical evolution}

Figure~\ref{fig:dynamics} presents the dynamical evolution of the thin shell driven by the ejecta. 
The model parameters are set to $\Gamma_\mathrm{max}=1.5$, $E_\mathrm{rel,49}\equiv E_\mathrm{rel}/(10^{49}\,\mathrm{erg})=3.5$, $A_\ast=120$, and $R_\mathrm{out}=350\,R_\odot$, which are later shown to reproduce the observed light curve of EP260321a. 

In the light-curve calculation, a photon emitted radially from the shell located at radius $R_\mathrm{sh}$ and epoch $t$ reaches the observer with a delay time of
$t-R_\mathrm{sh}/c$, relative to a photon emitted at the origin at $t=0$. 
Throughout this subsection, we therefore present the shell properties as functions of $t-R_\mathrm{sh}/c$, which serves as a useful proxy for the observer time.

The shell velocity $\beta_\mathrm{sh}$ gradually decreases as the shell sweeps up the surrounding CSM. The deceleration is initially efficient because the shell continuously accumulates mass from the dense environment. 
An important epoch corresponds to the emergence of the shell from the outer boundary of the dense CSM, namely when $R_\mathrm{sh}=R_\mathrm{out}$. 
Beyond this point, the model assumes that little additional CSM remains to decelerate the shell. 
Consequently, the shell transitions from a decelerating phase to an approximately coasting phase with a nearly constant velocity. For the fiducial model shown in Figure~\ref{fig:dynamics}, this transition occurs at 
$t-R_\mathrm{sh}/c\simeq10^3\,\mathrm{s}$, 
after which the shell velocity asymptotically approaches $\beta_\mathrm{sh}\simeq0.34$.

The shell optical depth, $\tau_\mathrm{sh}$, initially increases because mass is accumulated more rapidly than the shell expands. At later times, the continued expansion of the shell causes the optical depth to decrease. The shell eventually becomes optically thin, reaching $\tau_\mathrm{sh}=1$ at $t-R_\mathrm{sh}/c\simeq2\times 10^3\,\mathrm{s}$. 

The equilibrium temperature of the shell can be estimated from the shell internal energy density, $u_\mathrm{sh}$, assuming that the internal energy is dominated by radiation, 
$T_\mathrm{eq}=(u_\mathrm{sh}/a_\mathrm{r})^{1/4}$, where $a_\mathrm{r}$ being the radiation constant. 
This estimate further assumes that the shocked gas and radiation maintain local thermodynamic equilibrium (LTE). 
This LTE assumption is further examined later in Section \ref{sec:discussion}. 
Around the epoch at which the shell reaches the CSM outer radius, the equilibrium temperature decreases rapidly owing to the sharp decline in the radiation energy density retained within the shell.

\subsection{Synthetic light curves}
\begin{figure}[tbp]
\includegraphics[scale=0.75]{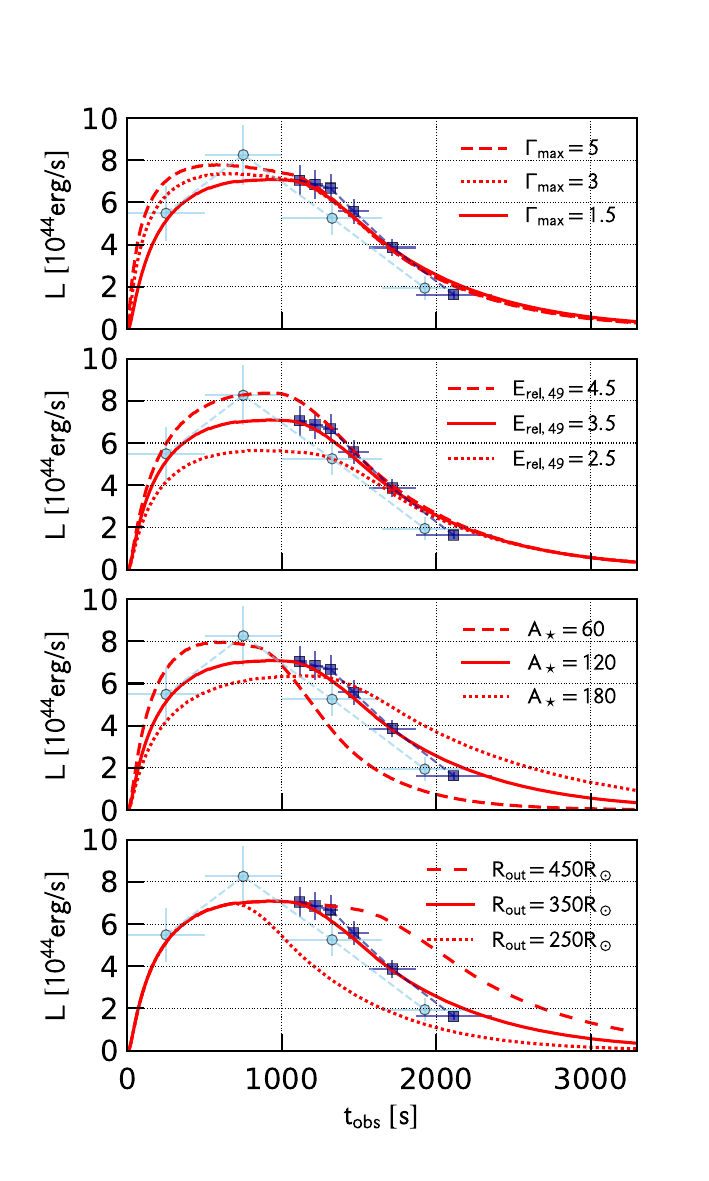}
\caption{Synthetic bolometric light curves compared with observations of EP260321a. 
In each panel, the bolometric luminosity inferred from WXT and FXT observations are plotted (circles and squares). 
The fiducial model is presented by a solid line, while models assuming different values of $\Gamma_\mathrm{max}$, $E_\mathrm{rel}$, $A_\ast$, and $R_\mathrm{out}$ (from top to bottom) are presented for highlighting the parameter dependence. }
\label{fig:lc_dependence}
\end{figure}

We now turn to the light-curve modeling of EP260321a.  
In the following, we take the model discussed in the previous section as a fiducial model. 

Our SBO model follows the evolution of frequency-integrated radiation and therefore predicts only the bolometric luminosity. 
In contrast, the Wide-field X-ray Telescope (WXT) and the Follow-up X-ray Telescope (FXT) onboard {\it Einstein Probe} observe over finite energy ranges of $0.5$--$4\,\mathrm{keV}$ and $0.3$--$10\,\mathrm{keV}$, respectively.

As discussed by \citet{2026arXiv260610014Y}, the X-ray spectra of EP260321a are well described by blackbody models with temperatures of $kT\simeq0.1$--$0.15\,\mathrm{keV}$. 
Assuming a Planck spectrum, the corresponding spectral peak lies at approximately $0.3$--$0.5\,\mathrm{keV}$, near the lower boundary of the WXT and FXT energy bands. 
We thus adopt the spectral fitting results presented by \citet{2026arXiv260610014Y} and convert the reported unabsorbed $0.4$--$2\,\mathrm{keV}$ fluxes to bolometric fluxes by assuming a Planck spectrum with the corresponding best-fit photon temperatures. This correction increases the inferred fluxes by approximately a factor of two. 

The WXT light curve exhibits a peak at $t_\mathrm{obs}\sim700\,\mathrm{s}$, although the precise peak time may remain uncertain because of the limited photon statistics. 
Observations with FXT commenced at $t_\mathrm{obs}\simeq1000\,\mathrm{s}$. Owing to its substantially larger effective area, FXT collected a much larger number of photons, enabling the construction of a high-quality light curve that clearly reveals the subsequent luminosity evolution. 

Figure~\ref{fig:lc_dependence} compares the observed WXT and FXT light curves with representative synthetic light curves from our SBO model. 
The presented synthetic light curves share similar rising and decaying characteristics, which have also been found in previous studies of wind breakout models \citep[e.g.,][]{2024ApJ...972..140K}. 
Overall, the synthetic light curves reproduce both the characteristic luminosity and duration. 
In particular, the models show remarkable agreement with the FXT observations during the declining phase, indicating that the SBO scenario in a dense CSM naturally accounts for the overall evolution of the prompt X-ray emission from EP260321a.

At later times, typically around $t_\mathrm{obs}\simeq1200$--$1300\,\mathrm{s}$, the fiducial light curve transitions to a steeper decline phase. 
This behavior reflects the emergence of the shocked shell from the CSM outer radius (Figure~\ref{fig:dynamics}). 
While the shell remains inside the CSM, a substantial fraction of the dissipated energy is efficiently converted into radiation and escapes through diffusion. 
Once the shell reaches the outer edge of the CSM, however, the internal energy supply suddenly stops, leading to a rapid decline in the observed luminosity.

\subsection{Parameter dependence}

We next investigate how each free parameter affects the resulting SBO light curve. Understanding these parameter dependencies is useful not only for interpreting the light-curve fitting results but also for clarifying the physical origin of the observed features in EP260321a.

\subsubsection{Maximum velocity and Ejecta energy}

The top panel of Figure~\ref{fig:lc_dependence} shows the model dependence on the maximum Lorentz factor; representative models with adjusted values of $E_\mathrm{rel}$ and $A_\ast$ for different values of the maximum Lorentz factor, $\Gamma_\mathrm{max}=5.0$, $3.0$, and $1.5$ are plotted. 
Although these models provide similarly good descriptions of the overall luminosity evolution, they exhibit systematically different rising behaviors.

Models with larger $\Gamma_\mathrm{max}$ display slightly more rapid rises toward peak luminosity. For the ejecta profile assumed in Equation~(\ref{eq:density}), extending the distribution to higher velocities places a larger fraction of the kinetic energy into the outermost, low-mass ejecta layers. These fast-moving layers encounter the dense CSM at earlier times and efficiently dissipate their kinetic energy through shock interaction. Consequently, the radiation output increases more rapidly, producing an earlier rise in the SBO light curve. 
As seen in Figure~\ref{fig:lc_dependence}, assuming $\Gamma_\mathrm{max}=1.5$, the light curve keeps rising even around $t_\mathrm{obs}\sim 500\,\mathrm{s}$, which is in agreement with the observed behavior. 

For a fixed ejecta velocity structure and CSM configuration, the kinetic energy of the high-velocity ejecta, $E_\mathrm{rel}$, primarily determines the overall energy budget available for radiation production. 
As shown in Figure~\ref{fig:lc_dependence} (the 2nd panel), increasing (decreasing) $E_\mathrm{rel}$ systematically shifts the light curve toward higher (lower) luminosities while preserving its overall shape and characteristic timescales.


\subsubsection{CSM density and outer radius\label{sec:csm_density}}

The CSM density parameter $A_\ast$ primarily governs the characteristic timescale of the SBO emission. As shown in the 3rd panel of Figure~\ref{fig:lc_dependence}, models with larger values of $A_\ast$ exhibit slower rises and longer burst durations.

In a denser CSM, the forward shock sweeps up more material, resulting in a more massive shocked shell. 
The increased shell mass enhances the optical depth and prolongs the diffusion timescale of photons trapped within the shell, thereby producing a more gradual rise toward peak luminosity. 
Furthermore, the shock is decelerated more efficiently in a denser CSM, delaying its emergence from the outer boundary of the dense medium. As a result, the SBO emission persists for a longer duration. 

The influence of the CSM outer radius $R_\mathrm{out}$ is illustrated in the bottom panel of Figure~\ref{fig:lc_dependence}. 
With the adopted parameter set, $R_\mathrm{out}$ determines when the ejecta–CSM interaction terminates. Once the shock emerges from the dense medium, the continuous conversion of ejecta kinetic energy into radiation effectively ceases (Figure~\ref{fig:dynamics}), causing the luminosity to decline rapidly. 
Earlier phases of the light curve remain largely unaffected because the interaction dynamics prior to reaching $R_\mathrm{sh}=R_\mathrm{out}$ are unchanged. 
As shown in Figure~\ref{fig:lc_dependence}, the model with $R_\mathrm{out}=450\,R_\odot$ exhibits more prolonged emission, whereas the model with $R_\mathrm{out}=250\,R_\odot$ undergoes a noticeably earlier decline. 

\subsection{Radiated energy versus burst duration}
\begin{figure*}[tbp]
\includegraphics[scale=0.6]{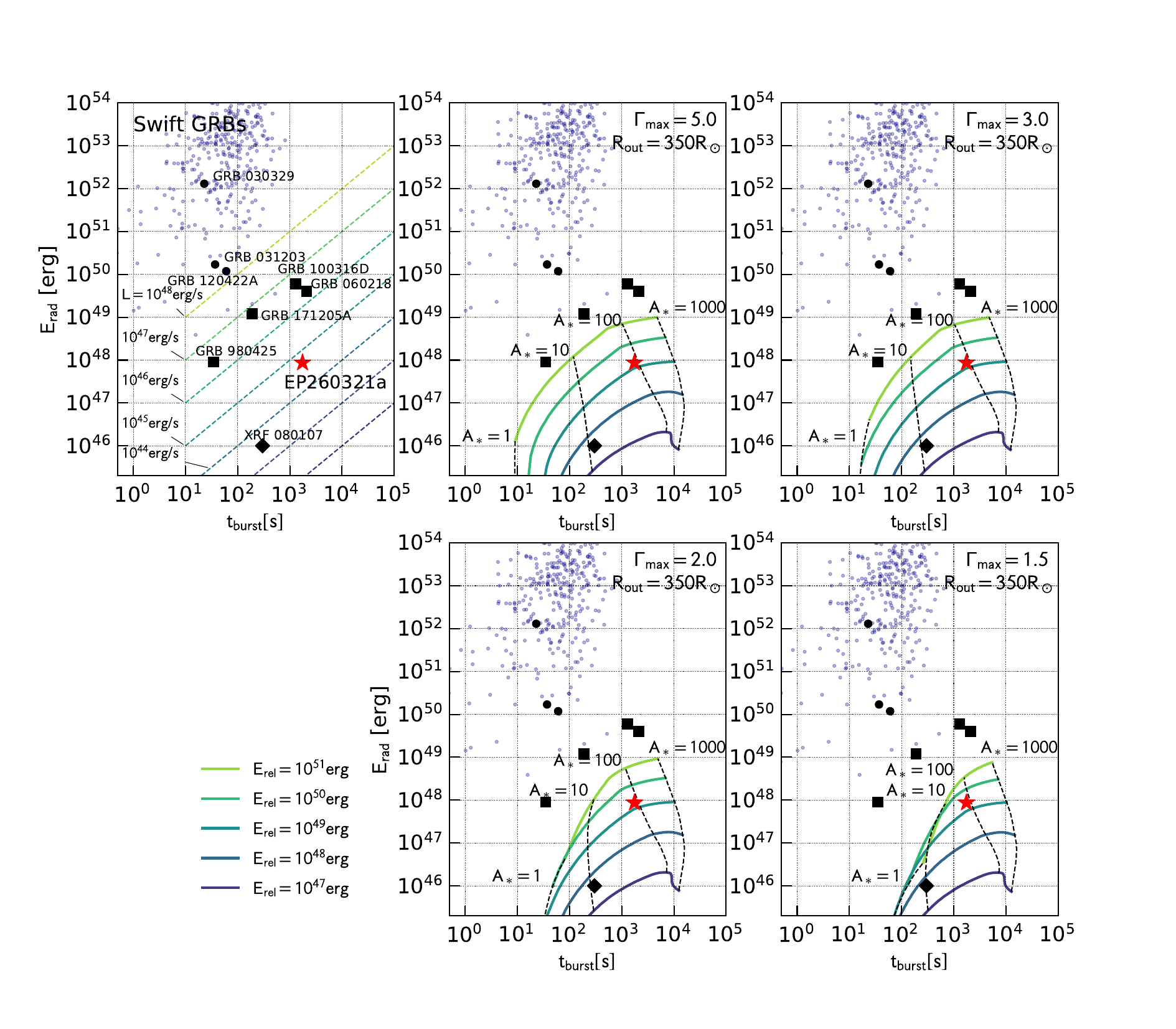}
\caption{Theoretical model grids in the radiated energy vs burst duration plane. The upper left panel shows the isotropic equivalent radiated energies and the burst duration for known GRBs and X-ray transients ($E_\mathrm{iso}$ and $T_{90}$ for GRBs), which include classical GRBs from {\it Swift} observations (dots) GRBs associated with SNe (circles), llGRBs (squares), and X-ray transients (diamond: XRF080107, star: EP260321a). 
The panel also includes lines with constant average luminosity $L=E_\mathrm{rad}/t_\mathrm{burst}$ from $10^{42}$ to $10^{48}\,\mathrm{erg}\,\mathrm{s}^{-1}$ (dashed lines). 
In the remaining panels, the predictions from the SBO model for different energy $E_\mathrm{rel}$ and the CSM density parameter $A_\ast$ are plotted. 
The models with $\Gamma_\mathrm{max}=5.0$, $3.0$, $2.0$, and $1.5$ while assuming a fixed $R_\mathrm{out}=350\,R_\odot$ are presented as indicated in the panels. 
In each panel, the models with fixed $E_\mathrm{rel}=10^{47}$, $10^{48}$, $10^{49}$, $10^{50}$, and $10^{51}\,\mathrm{erg}$, and varied $A_\ast$ from $1$ to $10^3$ are plotted as thick lines. 
}
\label{fig:diagram}
\end{figure*}

We further examine the global properties of SBO emission predicted by our model. 
Following the procedure adopted by \citet{2019ApJ...870...38S}, we calculate the radiated energy and the burst duration for a wide range of model parameters.

For each model light curve, the radiated energy $E_\mathrm{rad}$ is obtained by integrating the bolometric luminosity over the entire duration of the event. 
The burst duration $t_\mathrm{burst}$ is defined as the time interval over which the observer receives $90\%$ of the total radiated energy, namely from $t_\mathrm{obs}=0$ to $t_\mathrm{obs}=t_\mathrm{burst}$. 
This definition is analogous to the commonly used $T_{90}$ duration adopted for GRBs.

Figure~\ref{fig:diagram} shows the distributions of model predictions on the $E_\mathrm{rad}$--$t_\mathrm{burst}$ plane for four different values of the maximum Lorentz factor, $\Gamma_\mathrm{max}=5.0$, $3.0$, $2.0$, and $1.5$ and a fixed CSM outer radius $R_\mathrm{out}=350\,R_\odot$. 
The figure also includes observed high-energy transients reported in the literature, plotted using their isotropic-equivalent energies, $E_\mathrm{iso}$, and durations, $T_{90}$.

Although the observational quantities $E_\mathrm{iso}$ and $T_{90}$ are not strictly equivalent to the model quantities $E_\mathrm{rad}$ and $t_\mathrm{burst}$, such a comparison is useful for placing EP260321a in the broader context of fast high-energy transients. 
We note, however, that the measured duration of an event can depend on the observational energy band. 
Recent {\it Einstein Probe} observations have demonstrated that burst durations derived from soft X-ray observations can systematically differ from those measured in the gamma-ray regime \citep{2025ApJ...986..106G}. 
Therefore, caution is required when directly comparing classical GRBs with fast X-ray transients discovered by {\it Einstein Probe}.

Nevertheless, the $E_\mathrm{rad}$--$t_\mathrm{burst}$ diagram provides a convenient diagnostic for understanding how the observable properties of SBO emission depend on the underlying explosion and CSM parameters, and for assessing whether EP260321a occupies a region distinct from those of previously reported SBO candidates and llGRBs.

\subsubsection{Observed transient populations}

Figure~\ref{fig:diagram} (top-left panel) reveals a clear separation among the known populations of high-energy transients. 
Classical GRBs occupy the region characterized by the largest radiated energies and shortest durations. 
Most classical GRBs cluster around $E_\mathrm{iso}\gtrsim10^{51}$--$10^{53}\,\mathrm{erg}$ with $T_{90}$ ranging from a few seconds to several hundreds of seconds. 
The classical GRB population extends down to lower radiated energies of $E_\mathrm{iso}\sim10^{50}\,\mathrm{erg}$, where several events exhibiting properties intermediate between those of classical GRBs and llGRBs, such as GRB~031203 \citep{2004ApJ...609L...5M,2004Natur.430..648S} and GRB~120422A \citep{2012A&A...547A..82M,2014A&A...566A.102S}, have been identified.

Although this population spans a wide range of burst durations, from several tens to several thousands of seconds, their isotropic-equivalent luminosities remain relatively similar, typically around $10^{46}$--$10^{47}\,\mathrm{erg\,s^{-1}}$. 
This behavior suggests that llGRBs constitute a distinct population from classical GRBs, potentially reflecting differences in the explosion mechanism or the properties of the circumstellar environment.

With a reported (X-ray) radiated energy of $E_\mathrm{rad}=8.7\times10^{47}\,\mathrm{erg}$ and a burst duration of $t_\mathrm{burst}=1750\,\mathrm{s}$ \citep{2026arXiv260610014Y}, EP260321a occupies an even less energetic region of the diagram. 
Its duration is comparable to those of GRB~060218 and GRB~100316D, yet its radiated energy is nearly two orders of magnitude smaller than those of the two previously reported llGRBs. 
This difference clearly distinguishes EP260321a from the established llGRB population.

The X-ray transient associated with SN~2008D is located at even lower radiated energies, approximately two orders of magnitude below that of EP260321a. 
In this sense, EP260321a appears to bridge the gap between SN~2008D and the llGRB population, sharing the long durations characteristic of llGRBs while exhibiting a significantly lower radiated energy output. 
This intermediate location raises the possibility that EP260321a represents a transitional event linking ordinary SBOs and the more energetic transients associated with engine-driven explosions.

\subsubsection{Theoretical model grids}

The theoretical model grids shown in Figure~\ref{fig:diagram} illustrate how the SBO model can account for the diversity of observed X-ray and gamma-ray transients on the $E_\mathrm{rad}$--$t_\mathrm{burst}$ plane. 
In particular, the model parameters governing the high-velocity ejecta and the CSM influence the two observable quantities in qualitatively different ways.

For a fixed ejecta energy, increasing the CSM density results in longer burst durations and larger radiated energies. 
As we have discussed above (Figure~\ref{fig:lc_dependence}), this is caused by prolonged diffusion time and larger dissipated energy due to the denser CSM. 
Combining these two effects together, models with a fixed $E_\mathrm{rel}$ roughly follow a line with a constant average luminosity $E_\mathrm{rel}/t_\mathrm{burst}$=Const, except for high $A_\ast$ ($>100$) models. 
On the other hand, for a fixed $A_\ast$, increasing the kinetic energy contained in the high-velocity ejecta from $E_\mathrm{rel}=10^{47}\,\mathrm{erg}$ to $10^{51}\,\mathrm{erg}$ leads to a monotonic increase in the radiated energy. 
This trend primarily shifts the model predictions in the vertical direction on the $E_\mathrm{rad}$--$t_\mathrm{burst}$ diagram. 
These systematic trends are evident in all panels of Figure~\ref{fig:diagram}, irrespective of the adopted value of $\Gamma_\mathrm{max}$.

The location of EP260321a on the $E_\mathrm{rad}$--$t_\mathrm{burst}$ plane is well covered by the theoretical model grids, allowing a rough estimate of the physical parameters required to reproduce the observed event. 
Although the detailed shapes of the grids vary somewhat with $\Gamma_\mathrm{max}$, the regions consistent with EP260321a occupy a relatively narrow range in parameter space. 
In particular, the different model grids consistently suggest an ejecta energy of $E_{\mathrm{rel},49} \sim$ a few, together with a CSM density parameter of approximately $A_\ast \simeq 100$. 

On the other hand, llGRBs lie above the theoretical model grid even for $\Gamma_\mathrm{max}=5$ and $E_\mathrm{rel}=10^{51}\,\mathrm{erg}$. 
This indicates that reproducing these more energetic counterparts requires ejecta with larger energies and faster velocities. 
An energy deposition into material confined within a small solid angle, i.e., jet or bipolar explosion rather than a spherical point explosion, is a potential way to produce such high-velocity ejecta.

\section{Discussion\label{sec:discussion}}
In the following, we discuss the implications of the theoretical modeling for the SBO emission from EP260321a.

\subsection{Light-curve comparison with EP260321a}

Although the SBO model successfully reproduces the overall luminosity, duration, and declining behavior of the prompt X-ray emission from EP260321a, it may not perfectly reproduce the observed light curve. 
In particular, the observed luminosity appears to rise more sharply toward the peak than in the synthetic light curves presented in this work. 
This discrepancy is even more apparent when compared with the unabsorbed flux shown in Figure~1 of \citet{2026arXiv260610014Y}, which is derived directly from the observed count rates. 
By contrast, the declining phase observed by FXT is reproduced remarkably well by the models. 

The significance of this discrepancy remains uncertain because the WXT light curve around the peak suffers from relatively large statistical uncertainties. 
Nevertheless, if the delayed sharp peak is confirmed, it may indicate that additional physical ingredients not included in our fiducial model influence the SBO emission. 
Possible explanations include (1) a more complicated radial structure of the CSM and (2) departures from spherical symmetry in the ejecta and/or the CSM.

In the present study, we assume a wind-like CSM with a density profile of $\rho_\mathrm{w}\propto r^{-2}$, corresponding to steady mass loss at a constant rate and wind velocity. 
While this represents a physically motivated and commonly adopted prescription, it may be overly simplistic for describing the immediate environment of EP260321a/SN~2026gzf. 
The complex photometric variability inferred from the pre-explosion observations of SN~2026gzf \citep{2026arXiv260610009C} suggests that the progenitor experienced multiple episodes of enhanced mass loss prior to core collapse. 
Such activity could produce a CSM distribution substantially different from a smooth wind profile. 
For example, the dense material may instead be concentrated within one or more shells occupying relatively narrow radial ranges. 
Because the SBO light curve is sensitive to the density structure encountered by the shock, these more complicated CSM configurations may naturally modify the early luminosity evolution and delay the observed peak.
Exploring the full range of possible CSM structures is beyond the scope of the present work. Given the large number of additional degrees of freedom introduced by arbitrary radial density profiles, a systematic investigation of their effects on SBO emission is left for future studies.

Departures from spherical symmetry provide another possible explanation for the delayed peak \citep{2010ApJ...717L.154S,2011ApJ...727..104C,2013ApJ...779...60M,2018ApJ...856..146A,2021MNRAS.508.5766I,2024ApJ...976..147C,2026ApJ...999...56C}. 
Our previous two-dimensional radiation hydrodynamic simulations of bipolar shock breakout \citep{2016ApJ...825...92S} demonstrated that the resulting emission can peak substantially later than in corresponding spherical models, depending on the viewing angle. 
Although the physical setup considered in those simulations differs from that appropriate for EP260321a, the results illustrate the sensitivity of SBO light curves to multidimensional effects. 
The delayed peak observed in EP260321a may therefore provide an important clue to the geometry of the explosion and its circumstellar environment. 
Although the continuum polarization of SN~2026gzf at $16.5$ days suggests a rather spherical photosphere, the spectropolarimetric data indicate a globally axi-symmetric distribution of calcium above the photosphere \citep{2026arXiv260618881W}. 
This may imply a non-spherical explosion, affecting the SBO light curve.

Overall, the comparison with EP260321a suggests that the simple spherical wind-breakout model captures the dominant physical processes governing the X-ray emission, while the remaining discrepancies may encode valuable information regarding the progenitor's mass-loss history and the multidimensional nature of the explosion.

\subsection{X-ray spectra\label{sec:spectra}}

The X-ray spectra presented by \citet{2026arXiv260610014Y} are well described by blackbody models with photon temperatures of approximately $100$--$140\,\mathrm{eV}$. 
As already pointed out by \citet{2026arXiv260610014Y} and \citet{2026arXiv260610011R}, reproducing such temperatures within SBO models is challenging if complete thermal equilibrium between radiation and matter is assumed.

The observed peak X-ray luminosity,
$L_\mathrm{X,peak}\simeq8\times10^{44}\,\mathrm{erg\,s^{-1}}$,
measured at
$t_\mathrm{obs}=t_\mathrm{X,peak}\simeq700\,\mathrm{s}$,
together with the characteristic photon temperature, $k_\mathrm{B}T_\mathrm{ph}\simeq140\,\mathrm{eV}$ $(T_\mathrm{ph}
\simeq1.6\times10^6\,\mathrm{K})$ implies a blackbody radius of
\begin{equation}
\begin{aligned}
R_\mathrm{bb}
&=
\left(
\frac{L_\mathrm{X,peak}}
{4\pi\sigma_\mathrm{SB}T_\mathrm{ph}^4}
\right)^{1/2}
\\
&=
4\times10^{11}\,\mathrm{cm}
\left(
\frac{L_\mathrm{X,peak}}
{8\times10^{44}\,\mathrm{erg\,s^{-1}}}
\right)^{1/2}
\left(
\frac{T_\mathrm{ph}}
{140\,\mathrm{eV}}
\right)^{-2},
\end{aligned}
\end{equation}
where $\sigma_\mathrm{SB}$ is the Stefan--Boltzmann constant.

This inferred radius is substantially smaller than the shell radius predicted by our SBO models around the epoch of peak luminosity (Figure~\ref{fig:dynamics}). 
Furthermore, the corresponding expansion velocity,
$R_\mathrm{bb}/t_\mathrm{X,peak}
\simeq
6\times10^3\,\mathrm{km\,s^{-1}}$, is significantly lower than the Fe\,{\sc ii} velocities of approximately $(3$--$4)\times10^4\,\mathrm{km\,s^{-1}}$ measured in SN~2026gzf several days after explosion \citep{2026arXiv260610002M,2026arXiv260609992O,2026arXiv260610011R}. 
Under the assumption of homologous expansion, the photospheric radius inferred from the X-ray spectra therefore appears inconsistent with the characteristic ejecta velocities inferred from the optical observations. 

A small blackbody radius has also been claimed for SN~2008D \citep{2007MNRAS.375..240L,2008ApJ...683L.135C,2008cosp...37.3512X}, which indicates that deviation from LTE is likely common for SBO in dense wind \citep{2010ApJ...725..904N,2012ApJ...747...88N}. 
Therefore, understanding the spectral evolution of SBO emission warrants further studies of the treatment of radiative shocks and spectral formation in the shocked expanding gas \citep{2008PhRvL.100m1101L,2010ApJ...725...63B,2010ApJ...716..781K,2013MNRAS.429.3181T,2018MNRAS.474.2828I,2020MNRAS.499.4961I,2026MNRAS.547ag389I,2019MNRAS.484.3502I,2025MNRAS.541L..85I,2025MNRAS.543.2917I,2026PhRvD.113h3049A} in wide ranges of the breakout velocity and ambient density. 

The equilibrium temperatures directly obtained in our SBO calculations further emphasize this discrepancy. 
As shown in Figure~\ref{fig:dynamics}, the equilibrium temperature of the shocked shell remains below $10^6\,\mathrm{K}$ at $t-R_\mathrm{sh}/c\simeq10^2\text{--}10^3\,\mathrm{s}$, which corresponds to the epochs contributing dominantly to the observed X-ray emission. 
Consequently, the LTE temperatures predicted by the model are systematically lower than the temperatures inferred from the blackbody fits to the X-ray spectra.

In agreement with \citet{2026arXiv260610014Y} and \citet{2026arXiv260610011R}, these considerations suggest that a treatment assuming complete thermal equilibrium between radiation and matter is insufficient for describing the spectral evolution of EP260321a. 
Instead, non-LTE effects would play a crucial role in shaping the emergent X-ray luminosity and spectra, even when the bolometric properties remain unchanged. 

\subsection{CSM properties}
\subsubsection{CSM outer radius}
As discussed in Section~\ref{sec:csm_density}, the CSM density and outer radius primarily affect the rising and the declining timescales of the SBO light curve. 
Consequently, these parameters are relatively well constrained by the observed X-ray evolution. 
The light-curve modeling presented in Section~\ref{sec:result} suggests a dense CSM with $A_\ast\simeq120$ extending out to $R_\mathrm{out}\simeq 350\,R_\odot$. 
The former value is required for realizing a slow rise to the peak, while the latter value causes the luminosity to decline at $t_\mathrm{obs}\simeq 1200$--$1300\,\mathrm{s}$ (Figure~\ref{fig:lc_dependence}). 

The inferred CSM extent is broadly consistent with the SBO modeling of \citet{2026arXiv260610014Y}, who derived an outer radius of approximately $320\,R_\odot$. 
In contrast, \citet{2026arXiv260610011R}, based on the analytic framework developed by \citet{2021ApJ...910..128H}, suggested a more compact CSM with radii of $30$--$70\,R_\odot$ ($2$--$5\times10^{12}\,\mathrm{cm}$).

With the adopted parameters, the shock breakout happens when the shell is decelerated to $0.3$--$0.4c$, after which the shell travels at a coasting velocity. 
This velocity is higher than the Fe\,{\sc ii} absorption velocities of approximately $0.1c$ observed several days after the X-ray trigger \citep{2026arXiv260610002M,2026arXiv260609992O,2026arXiv260610011R}. 

As demonstrated in the case of SN~2017iuk associated with GRB~171205A, whose early spectra exhibited absorption features corresponding to ejecta velocities as high as $\sim0.3c$ \citep{2019Natur.565..324I}, detailed spectroscopic modeling of SN~2026gzf may reveal evidence for even faster ejecta components. 
Further investigations of the spectral evolution are therefore essential for clarifying the presence or absence of ejecta exceeding velocities of $\sim 0.1c$ several days after explosion \citep{2023MNRAS.522.2267M}.

\subsubsection{CSM mass}
The inferred CSM density parameter of $A_\ast\simeq120$ together with $R_\mathrm{out}=350\,R_\odot$ implies a CSM mass of $M_\mathrm{CSM}\sim 10^{-5}\,M_\odot$. 
The optically thick portion of the CSM enclosed within the photospheric radius $R_\mathrm{ph}\simeq1.2\times10^{13}\,\mathrm{cm}
\simeq170\,R_\odot$, contains $M_\mathrm{thick}\sim5\times 10^{-6}\,M_\odot$.

This estimated CSM mass is smaller than the shocked CSM mass of approximately $10^{-4}\,M_\odot$ obtained by \citet{2026arXiv260610014Y}. 
Their estimate was obtained by equating the radiated energy $E_\mathrm{rad}\simeq9\times10^{47}\,\mathrm{erg}$ to the kinetic energy of the shocked CSM, $M_\mathrm{CSM}v_\mathrm{sh}^2/2$, assuming a shock velocity of $v_\mathrm{sh}=0.1c$. 
This velocity is obtained for ejecta with a steep outer density profile, $\mathrm{d}\ln\rho/\mathrm{d}\ln r=-10$, characteristic of ordinary SN explosions \citep{1999ApJ...510..379M}. 

In contrast, our model adopts a shallower outer ejecta profile with $n=5$ (Equation~\ref{eq:density}), motivated by the presence of high-velocity ejecta in energetic SNe. 
In this case, the shell velocity during the SBO phase remains substantially higher, $\beta_\mathrm{sh}\simeq0.3\text{--}0.4$ (Figure~\ref{fig:dynamics}). 
Consequently, a smaller amount of CSM is sufficient for the reverse shock to dissipate the kinetic energy of the fastest ejecta and generate the observed SBO emission.

Therefore, the CSM mass required to explain EP260321a depends sensitively on the assumed structure of the outer ejecta and the breakout velocity. 
Optical spectra have been used to constrain the radial density structure of SN ejecta \citep[e.g.,][]{2000ApJ...545..407M}. 
For SN~2008D, the optical spectral modeling by \cite{2008Sci...321.1185M} assumes a density slope of $n=5.5$ above $17,000\,\mathrm{km}\,\mathrm{s}^{-1}$. 
For SN~2026gzf, \cite{2026arXiv260610009C} employ a density slope of $n=6$ for their modeling of spectra after maximum light. 
Another spectral model is provided by \cite{2026arXiv260618881W}, who suggest a density slope of $n=6.41$ to $6.14$ and a maximum velocity of $\sim 0.12c$. 
Determining whether SN~2026gzf indeed possessed a shallow, high-velocity outer envelope will require further observational constraints from early optical spectra combined with theoretical modeling.

\subsubsection{CSM formation scenario}

The inferred CSM density parameter of $A_\ast\simeq120$ corresponds to a mass-loss rate of
$\dot{M}\simeq1.2\times10^{-3}\,M_\odot\,\mathrm{yr}^{-1}$, assuming a wind velocity of $v_\mathrm{w}=10^3\,\mathrm{km\,s^{-1}}$. 
This value exceeds the typical mass loss rate for WR stars by more than two orders of magnitude. 
The CSM outer {radius probed by the X-ray SBO implies the onset time for the mass-loss episode. 
For $R_\mathrm{out}=350\,R_\odot$, the corresponding timescale is only $t_\mathrm{pre}\simeq3\,\mathrm{days}$ before the explosion. 
This timescale coincides with the silicon-burning phase expected during the final evolution of massive stars. 
The detection of pre-explosion optical variability \citep{2026arXiv260610009C} strongly suggests that the mechanism responsible for the photometric activity was also associated with the enhanced mass loss. 
Unfortunately, the available optical observations do not cover the period immediately preceding the explosion. 
Consequently, it remains unclear whether the precursor activity observed between $t=-11\,\mathrm{yr}$ and $-40\,\mathrm{days}$ \citep{2026arXiv260610009C} persisted up to the final days before core collapse, as would be required if the SBO-inferred CSM originated from a continuous pre-SN mass-loss episode.

Currently reported light-curve and spectral modeling of SN~2026gzf suggest a low ejecta mass ($\sim 1$--$2\,M_\odot$) combined with a high explosion energy exceeding several $10^{51}\,\mathrm{erg}$ (\citealt{2026arXiv260609992O,2026arXiv260610011R,2026arXiv260610014Y}, but see \citealt{2026arXiv260610002M}). 
To estimate the energetic requirement for producing the inferred CSM, we assume that the material originally resided in the outer envelope of a stripped progenitor with mass $M_\star=3\,M_\odot$ and radius $R_\star=1\,R_\odot$. The gravitational binding energy of an envelope mass $M_\mathrm{env}$ is then given by
\begin{equation}
\begin{aligned}
E_\mathrm{bind}
&=
\frac{GM_\star M_\mathrm{env}}{R_\star}
\\
&\simeq
10^{44}\,\mathrm{erg}
\left(
\frac{M_\star}{3\,M_\odot}
\right)
\left(
\frac{M_\mathrm{env}}{10^{-5}\,M_\odot}
\right)
\left(
\frac{R_\star}{1\,R_\odot}
\right)^{-1}.
\end{aligned}
\end{equation}
Depositing an energy comparable to this binding energy within a period of $t_\mathrm{pre}=R_\mathrm{out}/v_\mathrm{esc}$ with $v_\mathrm{esc}=(2GM_\star/R_\star)^{1/2}$ being the escape velocity, the required energy injection rate leads to
\begin{equation}
    \frac{E_\mathrm{bind}}{t_\mathrm{pre}}=
    1.4\times 10^{38}\,\mathrm{erg}\,\mathrm{s}^{-1}.
\end{equation}

The required binding energy remains substantially smaller than the energy expected to be transported to the envelope by wave heating during the oxygen- and silicon-burning phases, which may reach $10^{46}$--$10^{47}\,\mathrm{erg}$ \citep{2018MNRAS.476.1853F,2021ApJ...906....3W}. Therefore, if wave heating is indeed responsible for the pre-SN mass loss in EP260321a/SN~2026gzf, only a small fraction of the available wave energy can be converted into the kinetic energy of escaping material.
In this sense, the relatively modest CSM mass inferred from the SBO modeling may itself provide an important constraint on the efficiency of wave-driven mass loss in stripped-envelope progenitors. Clarifying whether such low-efficiency mass ejection naturally emerges from stellar evolution models warrants future theoretical studies. 

An important caveat to the above discussion is the potential role of observational selection effects. 
The CSM properties inferred from EP260321a unlikely represent the full population of pre-supernova mass-loss events. 
Instead, X-ray surveys are naturally biased toward detecting SBO emission from progenitor systems whose CSM densities place the spectral peak in the X-ray band. 
If the progenitor had undergone substantially more intense mass loss, the explosion would have occurred in a denser environment, likely producing a more luminous SBO signal with a spectral peak shifted toward the ultraviolet. 
Therefore, while the mass-loss histories of stripped-envelope supernova progenitors may exhibit considerable diversity, X-ray-selected SBO events may preferentially sample systems containing only modest amounts of circumstellar material. 
A more complete picture of the SBO population will be uncovered with forthcoming ultraviolet transient surveys such as {\it ULTRASAT} \citep{2024ApJ...964...74S} and the {\it Ultraviolet Explorer} \citep{2021arXiv211115608K}.

\section{Summary\label{sec:summary}}
In this work, we applied our SBO model to the newly discovered X-ray transient EP260321a/SN~2026gzf and investigated whether its prompt X-ray emission can be explained as an SBO occurring within a dense CSM. 
Our main findings are summarized as follows.

\begin{enumerate}
\item{} The wind SBO scenario, in which the shock driven by SN ejecta emerges from the photosphere embedded within a dense CSM, successfully reproduces the overall luminosity, duration, and declining behavior of the X-ray light curve of EP260321a.
\item{} The observed X-ray emission requires a dense wind characterized by a mass-loss rate $\dot{M}\simeq1.2\times10^{-3}\,M_\odot\,\mathrm{yr}^{-1}$
for a wind velocity of $10^3\,\mathrm{km\,s^{-1}}$. The CSM must extend to $R_\mathrm{out}\simeq 350\,R_\odot$, implying a CSM mass of $M_\mathrm{CSM}\sim 10^{-5}\,M_\odot$. 
\item{} The SBO emission is powered by high-velocity ejecta with a kinetic energy of approximately
$E_\mathrm{rel}\simeq 3.5\times 10^{49}\,\mathrm{erg}$, assuming a maximum Lorentz factor of $\Gamma_\mathrm{max}=1.5$, (corresponding to the initial maximum velocity of $\sim0.7c$).
\item{} Although the bolometric light curve is well reproduced, the observed X-ray spectra remain difficult to explain under the assumption of complete thermal equilibrium between radiation and matter. This indicates that non-LTE effects play an important role in shaping the emergent spectra of SBOs occurring in dense CSM environments.
\item{} The inferred ejecta energy and CSM mass depend on the assumed structure of the high-velocity ejecta. In this work, we adopted an outer density profile with a slope of $n=5$, motivated by observations and multidimensional simulations of energetic explosions. 
However, alternative ejecta density structures cannot be ruled out. 
In particular, detailed optical spectral modeling can provide independent constraints on the outer density slope. 
\end{enumerate}

These results support the emerging picture that at least a subset of highly energetic stripped-envelope supernovae explode within dense CSM created shortly before core collapse. The inferred CSM masses of $\sim 10^{-5}M_\odot$ suggest that only modest amounts of material are required to produce observable SBO signatures, potentially offering a sensitive probe of pre-SN mass-loss processes operating during the final stages of stellar evolution, as well as the outer structure of the high-velocity ejecta. 

The advent of wide-field X-ray survey missions, such as \textit{Einstein Probe} and \textit{SVOM}, is opening a new observational window onto these rare transients. As the sample of SBO candidates grows, combining prompt X-ray observations with extensive multi-wavelength follow-up will provide increasingly stringent constraints on the properties of the progenitor stars, the origin of dense CSM, and the physical mechanisms driving the most energetic CCSN explosions.

\begin{acknowledgments}
A.S. acknowledges support by Japan Society for the Promotion of Science (JSPS) KAKENHI Grant Number JP22K03690 and JP26H00823. 
K.M. is supported by JSPS KAKENHI Grant JP24KK0070
JP23H04894. 
T.S. is supported by JSPS KAKENHI Grant JP22K03671. 
\end{acknowledgments}

\begin{contribution}
The authors thank the referee for carefully reading the manuscript and providing helpful suggestions. 
A.S. conceived and led the project, performed the calculations, analyzed the results, and wrote the manuscript. 
K.M. and T.S. contributed through scientific discussions, interpretation of the results, and comments on the manuscript. 
All authors reviewed and approved the final manuscript.


\end{contribution}

%

\software{NumPy \citep{harris2020array}, 
          Matplotlib \citep{Hunter:2007}}


\appendix



\bibliography{sample702}{}
\bibliographystyle{aasjournalv7.1}



\end{document}